\documentclass{article}
\usepackage{fullpage}
\usepackage[american]{babel}
\usepackage{amssymb}
\usepackage{algorithmic}
\usepackage{algorithm}

\newcommand{\eqn}[1]{(\ref{#1})}
\newcommand{\beql}[1]{\begin{equation}\label{#1}}
\newcommand{\eeq}{\end{equation}}
\newtheorem{theo}{Theorem}

\title{A Dynamic Programming Solution to a Generalized LCS Problem}
\author{Lei Wang,Xiaodong Wang,Yingjie Wu, and Daxin Zhu}

\begin{document}
\maketitle

\begin{abstract}
In this paper, we consider a generalized longest common subsequence problem, the string-excluding constrained LCS problem. For the two input sequences $X$ and $Y$ of lengths $n$ and $m$, and a constraint string $P$ of length $r$, the problem is to find a common subsequence $Z$ of $X$ and $Y$ excluding $P$ as a substring and the length of $Z$ is maximized. The problem and its solution were first proposed by  Chen  and  Chao\cite{1}, but we found that their algorithm can not solve the problem correctly.
A new dynamic programming solution for the STR-EC-LCS problem is then presented in this paper. The correctness of the new algorithm is proved. The time complexity of the new algorithm is $O(nmr)$.
\end{abstract}

\section{Introduction}
In this paper, we consider a generalized longest common subsequence problem. The longest common subsequence (LCS) problem is a well-known measurement for computing the similarity of two strings. It can be widely applied in diverse areas, such as file comparison, pattern matching and computational biology\cite{2,5,6,8,9}.

A sequence is a string of characters over an alphabet $\sum$. A subsequence of a sequence $X$ is obtained by deleting zero or more characters from $X$ (not necessarily contiguous). A substring of a sequence $X$ is a subsequence of successive characters within $X$.

For a given sequence $X=x_1x_2\cdots x_n$ of length $n$, the $i$th character of $X$ is denoted as $x_i \in \sum$ for any $i=1,\cdots,n$. A substring of $X$ from position $i$ to $j$ can be denoted as $X[i:j]=x_ix_{i+1}\cdots x_j$. A substring $X[i:j]=x_ix_{i+1}\cdots x_j$ is called a prefix or a suffix of $X$ if $i=1$ or $j=n$, respectively.

Given two sequences $X$ and $Y$, the longest common subsequence (LCS) problem is to find a subsequence of $X$ and $Y$ whose length is the longest among all common subsequences of the two given sequences.

For some biological applications some constraints must be applied to the LCS problem. These kinds of variant of the LCS problem are called the constrained LCS (CLCS) problem\cite{10}. Recently, Chen and Chao\cite{1} proposed the more generalized forms of the CLCS problem, the generalized constrained longest common subsequence (GC-LCS) problem.
For the two input sequences $X$ and $Y$ of lengths $n$ and $m$,respectively, and a constraint string $P$ of length $r$, the GC-LCS problem is a set of four problems which are to find the LCS of $X$ and $Y$ including/excluding $P$ as a subsequence/substring, respectively. The four generalized constrained LCS can be summarized in Table 1.

\begin{table}[ht]
\begin{center}
\caption{The GC-LCS problems}
\begin{tabular}{lll}
\hline\hline
Problem & Input & Output\\\hline
SEQ-IC-LCS &$X$,$Y$, and $P$ & The longest common subsequence of $X$ and $Y$ including $P$ as a subsequence\\
STR-IC-LCS &$X$,$Y$, and $P$ & The longest common subsequence of $X$ and $Y$ including $P$ as a substring\\
SEQ-EC-LCS &$X$,$Y$, and $P$ & The longest common subsequence of $X$ and $Y$ excluding $P$ as a subsequence\\
STR-EC-LCS &$X$,$Y$, and $P$ & The longest common subsequence of $X$ and $Y$ excluding $P$ as a substring\\
\hline
\end{tabular}
\end{center}
\end{table}

We will discuss the STR-EC-LCS problem in this paper. We have noticed that a previous proposed dynamic programming algorithm for the STR-EC-LCS problem\cite{1} can not correctly solve the problem. A new dynamic solution for the STR-EC-LCS problem is then presented in this paper. The correctness of the new algorithm is proved. The time complexity of the new algorithm is $O(nmr)$.

The organization of the paper is as follows.

In the following 4 sections we describe our presented dynamic programming algorithm for the STR-EC-LCS problem.

In section 2 we review the dynamic programming algorithm for the STR-EC-LCS problem proposed by Chen and Chao\cite{1}.
We point out that their algorithm will not work for a simple counterexample.
In section 3 we give a new dynamic solution for the STR-EC-LCS problem with time complexity $O(nmr)$ in a different point of view.
In section 4 we discuss the issues to implement the algorithm efficiently.
Some concluding remarks are in section 5.

\section{A Proposed Dynamic Programming Algorithm}
In this section, we will focus on the STR-EC-LCS problem and its solution proposed previously by Chen and Chao\cite{1}. As noted in table 1, for the two input sequences $X$ and $Y$ of lengths $n$ and $m$, and a constraint string $P$ of length $r$, the STR-EC-LCS problem is to find an LCS $Z$ of $X$ and $Y$ excluding $P$ as a substring.

Let $L(i,j,k)$ denote the length of an LCS of $X[1:i]$ and $Y[1:j]$ excluding $P[1:k]$ as a substring. Chen and Chao gave a recursive formula (1) for computing $L(i,j,k)$ as follows.

\beql{eq21}
L(i,j,k)=\left\{\begin{array}{ll} L(i-1,j-1,k) & \texttt{if } k=1 \texttt{ and } x_i=y_j=p_k, \\
1+\max \{L(i-1,j-1,k-1),L(i-1,j-1,k)\} & \texttt{if } k\geq 2 \texttt{ and } x_i=y_j=p_k, \\
1+L(i-1,j-1,k)& \texttt{if } x_i=y_j \texttt{ and } (k=0, \texttt{ or  } k>0 \texttt{ and } x_i\neq p_k), \\
\max\left\{ L(i-1,j,k),L(i,j-1,k) \right\} & \texttt{if } x_i\neq y_j.
\end{array} \right.
\eeq

The boundary conditions of this recursive formula are $L(i,0,k) = L(0,j,k) = 0$ for any $0\leq i\leq n, 0\leq j\leq m$, and $0\leq k \leq r$.

The correctness of the recursive formula (1) was based on Theorem 3 of their paper\cite{1} as follows.

\begin{theo}\label{th3}
\verb"(Chen and Chao 2011)"
Let $S_{i,j,k}$ denote the set of all LCSs of $X[1:i]$ and $Y[1:j]$ excluding $P[1:k]$ as a substring. If $Z[1:l]\in S_{i,j,k}$, the following conditions hold:

\renewcommand\labelenumi{(\theenumi)}
\begin{enumerate}

\item If $x_i=y_j=p_k$ and $k = 1$, then $z_l\neq x_i$ and $Z[1:l]\in S_{i-1,j-1,k}$.

\item If $x_i=y_j=p_k$ and $k\geq 2$, then $z_l=x_i=y_j=p_k$ and $z_{l-1}=p_{k-1}$ implies $Z[1:l-1]\in S_{i-1,j-1,k-1}$.

\item If $x_i=y_j=p_k$ and $k\geq 2$, then $z_l=x_i=y_j=p_k$ and $z_{l-1}\neq p_{k-1}$ implies $Z[1:l-1]\in S_{i-1,j-1,k}$.

\item If $x_i=y_j=p_k$ and $k\geq 2$, then $z_l\neq x_i$ implies $Z[1:l]\in S_{i-1,j-1,k}$.

\item If If $x_i=y_j$ and $x_i\neq p_k$, then $z_l=x_i=y_j$ and $Z[1:l-1]\in S_{i-1,j-1,k}$.

\item If $x_i\neq y_j$, then $z_l\neq x_i$ implies $Z[1:l]\in S_{i-1,j,k}$.

\item If $x_i\neq y_j$, then $z_l\neq y_j$ implies $Z[1:l]\in S_{i,j-1,k}$.
\end{enumerate}

\end{theo}

Since a common subsequence of $X[1:i]$ and $Y[1:j]$ excluding $P[1:k-1]$ as a substring is also a common subsequence of $X[1:i]$ and $Y[1:j]$ excluding $P[1:k]$ as a substring, by the definition of $L(i,j,k)$, we know that $L(i,j,k)\geq L(i,j,k-1)$ is always true.
Therefore, the recursive formula (1) can be further reduced to the recursive formula (2).
\beql{eq22}
L(i,j,k)=\left\{\begin{array}{ll} L(i-1,j-1,k) & \texttt{if } k=1 \texttt{ and } x_i=y_j=p_k, \\
1+L(i-1,j-1,k) & \texttt{if } k\geq 2 \texttt{ and } x_i=y_j=p_k, \\
1+L(i-1,j-1,k)& \texttt{if } x_i=y_j \texttt{ and } (k=0, \texttt{ or  } k>0 \texttt{ and } x_i\neq p_k), \\
\max\left\{ L(i-1,j,k),L(i,j-1,k) \right\} & \texttt{if } x_i\neq y_j.
\end{array} \right.
\eeq
Furthermore, the most important thing is that the above Theorem was only stated but without a strict proof. Therefore, the correctness of the proposed algorithm can not be guaranteed. For example, if $X=abbb, Y=aab$ and $P=ab$, the values of $L(i,j,k), 1\leq i\leq 4,1\leq j\leq 3,0\leq k\leq 2$ computed by recursive formula (1) and (2) are listed in Table 2.

\begin{table}[ht]
\caption{$L(i,j,k)$ computed by recursive formula (1) and (2)}
\begin{center}
\begin{tabular}{|c|ccc|ccc|ccc|}
\hline
&&$k=0$&&&$k=1$&&&$k=2$&\\\hline
$i=1$&1&1&1&0&0&0&1&1&1\\
$i=2$&1&1&2&0&0&1&1&1&2\\
$i=3$&1&1&2&0&0&1&1&1&2\\
$i=4$&1&1&2&0&0&1&1&1&2\\
\hline
\end{tabular}
\end{center}
\end{table}

From Table 2 we know that the final answer is $L(4,3,2)=2$ which is computed by the formula that $L(4,3,2)=1+L(3,2,2)$ since in this case $k\geq 2$ and $a_4=b_3=p_2='b'$. But, this is a wrong answer, since the correct answer should be 1.

We have tried to modify the recursive formula (1) or (2) to a correct one, but failed.

In next section, we will investigate the problem in a different way and finally present a correct dynamic solution for the STR-EC-LCS problem.

\section{Our New Dynamic Programming Solution}

For the two input sequences $X=x_1x_2\cdots x_n$ and $Y=y_1y_2\cdots y_m$ of lengths $n$ and $m$, respectively, and a constraint string $P=p_1p_2\cdots p_r$ of length $r$, we want to find an LCS of $X$ and $Y$ excluding $P$ as a substring.

In the description of our new algorithm, a function $\sigma$ will be mentioned frequently. For any string $S$ and a fixed constraint string $P$, the length of the longest suffix of $S$ that is also a prefix of $P$ is denoted by function $\sigma(S)$.

The symbol $\oplus$ is also used to denote the string concatenation.

For example, if $P=aaba$ and $S=aabaaab$, then substring $aab$ is the longest suffix of $S$ that is also a prefix of $P$, and therefore $\sigma(S)=3$.

It is readily seen that  $S\oplus P=aabaaabaaba$.

Let $Z(i,j,k)$ denote the set of all LCSs of $X[1:i]$ and $Y[1:j]$ excluding $P$ as a substring and $\sigma(z)=k$ for each $z\in Z(i,j,k)$.
The length of of an LCS in $Z(i,j,k)$ is denoted as $f(i,j,k)$.

If we can compute $f(i,j,k)$ for any $1\leq i\leq n, 1\leq j\leq m$, and $0\leq k<r$ efficiently, then the length of an LCS of $X$ and $Y$ excluding $P$ as a substring must be $\max\limits_{0\leq t<r}\left\{f(n,m,t)\right\}$.

We can give a recursive formula for computing $f(i,j,k)$ by following Theorem.

\begin{theo}\label{th1}
For the two input sequences $X=x_1x_2\cdots x_n$ and $Y=y_1y_2\cdots y_m$ of lengths $n$ and $m$, respectively, and a constraint string $P=p_1p_2\cdots p_r$ of length $r$, let $Z(i,j,k)$ denote the set of all LCSs of $X[1:i]$ and $Y[1:j]$ excluding $P$ as a substring and $\sigma(z)=k$ for each $z\in Z(i,j,k)$.

The length of of an LCS in $Z(i,j,k)$ is denoted as $f(i,j,k)$.

For any $1\leq i\leq n, 1\leq j\leq m$, and $0\leq k < r$, $f(i,j,k)$ can be computed by the following recursive formula \eqn{eq31}.

\beql{eq31}
f(i,j,k)=\left\{\begin{array}{ll}
\max\left\{ f(i-1,j,k),f(i,j-1,k) \right\} & \texttt{if } x_i\neq y_j,\\
\max\left\{
f(i-1,j-1,k),1+\max\limits_{0\leq t<r}\left\{f(i-1,j-1,t)|\sigma(P[1:t]\oplus x_i)=k\right\}
\right\} & \texttt{if } x_i= y_j.
\end{array} \right.
\eeq

The boundary conditions of this recursive formula are $f(i,0,k) = f(0,j,k) = 0$ for any $0\leq i\leq n, 0\leq j\leq m$, and $0\leq k \leq r$.
\end{theo}

\noindent{\bf Proof.}

For any $1\leq i\leq n, 1\leq j\leq m$, and $0\leq k < r$, suppose $f(i,j,k) = t$ and $z=z_1,\cdots, z_t\in Z(i,j,k)$.

First of all, we notice that for each pair $(i',j'), 1\leq i'\leq n, 1\leq j'\leq m$,such that $i'\leq i$ and $j'\leq j$, we have $f(i',j',k) \leq f(i,j,k)$, since a common subsequence $z$ of $X[1:i']$ and $Y[1:j']$ excluding $P$ as a substring and $\sigma(z)=k$ is also a common subsequence of $X[1:i]$ and $Y[1:j]$ excluding $P$ as a substring and $\sigma(z)=k$.

(1) In the case of $x_i\neq y_j$, we have $x_i\neq z_t$ or $y_j\neq z_t$.

(1.1)If $x_i\neq z_t$, then $z=z_1,\cdots, z_t$ is a common subsequence of $X[1:i-1]$ and $Y[1:j]$ excluding $P$ as a substring and $\sigma(z_1,\cdots, z_t)=k$, and so $f(i-1,j,k) \geq t$. On the other hand, $f(i-1,j,k)\leq f(i,j,k) = t$. Therefore, in this case we have $f(i,j,k) = f(i-1,j,k)$.

(1.2)If $y_j\neq z_t$, then we can prove similarly that in this case, $f(i,j,k) = f(i,j-1,k)$.

Combining the two subcases we conclude that in the case of $x_i\neq y_j$, we have $f(i,j,k)=\max\left\{ f(i-1,j,k),f(i,j-1,k) \right\}$.

(2) In the case of $x_i=y_j$, there are also two cases to be distinguished.

(2.1)If $x_i=y_j\neq z_t$,  then $z=z_1,\cdots, z_t$ is also a common subsequence of $X[1:i-1]$ and $Y[1:j-1]$ excluding $P$ as a substring and $\sigma(z_1,\cdots, z_t)=k$, and so $f(i-1,j-1,k) \geq t$. On the other hand, $f(i-1,j-1,k)\leq f(i,j,k) = t$. Therefore, in this case we have $f(i,j,k) = f(i-1,j-1,k)$.

(2.2)If $x_i=y_j=z_t$, then $f(i,j,k) = t>0$ and $z=z_1,\cdots, z_t$ is an LCS of $X[1:i]$ and $Y[1:j]$ excluding $P$ as a substring and $\sigma(z_1,\cdots, z_t)=k$, and thus $z_1,\cdots, z_{t-1}$ is a common subsequence of $X[1:i-1]$ and $Y[1:j-1]$ excluding $P$ as a substring.

Let $\sigma(z_1,\cdots, z_{t-1})=q$ and $f(i-1,j-1,q)=s$.
Then $z_1,\cdots, z_{t-1}$ is a common subsequence of $X[1:i-1]$ and $Y[1:j-1]$ excluding $P$ as a substring and $\sigma(z_1,\cdots, z_{t-1})=q$.
Therefore, we have

\beql{eq32}
f(i-1,j-1,q)=s\geq t-1.
\eeq

Let $v=v_1,\cdots, v_s\in Z(i-1,j-1,q)$ is an LCS of $X[1:i-1]$ and $Y[1:j-1]$ excluding $P$ as a substring and $\sigma(v_1,\cdots, v_s)=q$. Then
$\sigma((v_1,\cdots, v_s)\oplus x_i)=\sigma(P[1:q]\oplus x_i)=k$, and thus $(v_1,\cdots, v_s)\oplus x_i$ is a common subsequence of $X[1:i]$ and $Y[1:j]$ excluding $P$ as a substring and $\sigma((v_1,\cdots, v_s)\oplus x_i)=k$.

Therefore,
\beql{eq33}
f(i,j,k)=t\geq s+1.
\eeq

Combining \eqn{eq32} and \eqn{eq33} we have $s=t-1$. Therefore, $z_1,\cdots, z_{t-1}$ is an LCS of $X[1:i-1]$ and $Y[1:j-1]$ excluding $P$ as a substring and $\sigma(z_1,\cdots, z_{t-1})=q$.

In other words,
\beql{eq34}
f(i,j,k)\leq 1+\max\limits_{0\leq q<r}\left\{f(i-1,j-1,q)|\sigma(P[1:q]\oplus x_i)=k\right\}
\eeq

On the other hand, for any $0\leq q<r$, if $f(i-1,j-1,q)=s$ and $\sigma(P[1:q]\oplus x_i)=k$, then for any $v=v_1,\cdots, v_s\in Z(i-1,j-1,q)$, $v\oplus x_i$ is a common subsequence of $X[1:i]$ and $Y[1:j]$ and $\sigma(v\oplus x_i)=k$. Since $v$ excludes $P$ as a substring and $\sigma(v\oplus x_i)=k<r$, $v\oplus x_i$ is a common subsequence of $X[1:i]$ and $Y[1:j]$ excluding $P$ as a substring. Furthermore, $v\oplus x_i$ is a common subsequence of $X[1:i]$ and $Y[1:j]$ excluding $P$ as a substring and $\sigma(v\oplus x_i)=k$. Therefore, $f(i,j,k)=t\geq 1+s=1+f(i-1,j-1,q)$, and so we conclude that,
\beql{eq35}
f(i,j,k)\geq 1+\max\limits_{0\leq q<r}\left\{f(i-1,j-1,q)|\sigma(P[1:q]\oplus x_i)=k\right\}
\eeq

Combining \eqn{eq34} and \eqn{eq35} we have, in this case,
\beql{eq36}
f(i,j,k)= 1+\max\limits_{0\leq q<r}\left\{f(i-1,j-1,q)|\sigma(P[1:q]\oplus x_i)=k\right\}
\eeq

Combining the two subcases in the case of $x_i=y_j$, we conclude that the recursive formula \eqn{eq31} is correct for the case $x_i=y_j$.

The proof is complete. 
\hfill $\blacksquare$

\section{The Implementation of the Algorithm}
According to Theorem \ref{th1}, our new algorithm for computing $f(i,j,k)$ is a standard 2-dimensional dynamic programming algorithm. By the recursive formula \eqn{eq31}, the new dynamic programming algorithm for computing $f(i,j,k)$ can be implemented as the following Algorithm 1.
\begin{algorithm}
\caption{STR-EC-LCS}
{\bf Input:} Strings $X=x_1\cdots x_n$, $Y=y_1\cdots y_m$ of lengths $n$ and $m$, respectively, and a constraint string $P=p_1\cdots p_r$ of lengths $r$\\
{\bf Output:} The length of an LCS of $X$ and $Y$ excluding $P$ as a substring
\begin{algorithmic}[1]
\FORALL{$i,j,k$ , $0\leq i\leq n, 0\leq j\leq m$, and $0\leq k \leq r$}
\STATE $f(i,0,k) \leftarrow 0, f(0,j,k) \leftarrow 0$ \{boundary condition\}
\ENDFOR
\FOR{$i=1$ to $n$}
\FOR{$j=1$ to $m$}
\FOR{$k=0$ to $r$}
\IF {$x_i\neq y_j$}
\STATE $f(i,j,k) \leftarrow \max\{f(i-1,j,k),f(i,j-1,k)\}$
\ELSE
\STATE $u \leftarrow \max\limits_{0\leq t<r}\left\{f(i-1,j-1,t)|\sigma(P[1:t]\oplus x_i)=k\right\}$
\STATE $f(i,j,k) \leftarrow \max\{f(i-1,j-1,k),1+u\}$
\ENDIF
\ENDFOR
\ENDFOR
\ENDFOR
\RETURN $\max\limits_{0\leq t<r}\{f(n,m,t)\}$
\end{algorithmic}
\end{algorithm}

To implement our new algorithm efficiently, the most important thing is to compte $\sigma(P[1:k]\oplus x_i)$ for each $0\leq k<r$ and $x_i, 1\leq i\leq n$, in line 10 efficiently.

It is obvious that $\sigma(P[1:k]\oplus x_i)=k+1$ for the case of $x_i=p_{k+1}$. It will be more complex to compute $\sigma(P[1:k]\oplus x_i)$ for the case of $x_i\neq p_{k+1}$. In this case the length of matched prefix of $P$ has to be shortened to the largest $t<k$ such that $p_{k-t+1}\cdots p_k=p_1\cdots p_t$ and $x_i=p_{t+1}$. Therefore, in this case, $\sigma(P[1:k]\oplus x_i)=t+1$.

This computation is very similar to the computation of the prefix function in KMP algorithm for solving the string matching problem\cite{3,7}.

For a given string $S=s_1\cdots s_n$, the prefix function $kmp(i)$ denotes the length of the longest prefix of $s_1\cdots s_{i-1}$ that matches a suffix of $s_1\cdots s_i$. For example, if $S=ababaa$, then $kmp(1),\cdots, kmp(6)=0,0,1,2,3,1$.

For the constraint string $P=p_1\cdots p_r$ of lengths $r$, its prefix function $kmp$ can be pre-computed in $O(r)$ time as follows.

\begin{algorithm}
\caption{Prefix Function}
{\bf Input:} String $P=p_1\cdots p_r$\\
{\bf Output:} The prefix function $kmp$ of $P$
\begin{algorithmic}[1]
\STATE $kmp(0) \leftarrow -1$
\FOR{$i=2$ to $r$}
\STATE $k \leftarrow 0$\\
\WHILE{$k\geq 0 \ \AND\  p_{k+1}\neq p_i$}
\STATE $k \leftarrow kmp(k)$\\
\ENDWHILE
\STATE $k \leftarrow k+1$\\
\STATE $kmp(i) \leftarrow k$\\
\ENDFOR
\end{algorithmic}
\end{algorithm}

With this pre-computed prefix function $kmp$, the function $\sigma(P[1:k]\oplus ch)$ for each character $ch\in\sum$ and $1\leq k\leq r$ can be described as follows.

\begin{algorithm}
\caption{$\sigma(k,ch)$}
{\bf Input:} String $P=p_1\cdots p_r$, integer $k$ and character $ch$\\
{\bf Output:} $\sigma(P[1:k]\oplus ch)$
\begin{algorithmic}[1]
\WHILE{$k\geq 0 \ \AND\  p_{k+1}\neq ch$}
\STATE $k \leftarrow kmp(k)$\\
\ENDWHILE
\RETURN $k+1$
\end{algorithmic}
\end{algorithm}

Then, we can compute an index $t^*$ such that $$f(i-1,j-1,t^*)=\max\limits_{0\leq t<r}\left\{f(i-1,j-1,t)|\sigma(P[1:t]\oplus x_i)=k\right\}$$ in line 10 of Algorithm 1 by the following Algorithm 4.

\begin{algorithm}
\caption{$\max\sigma(i,j,k)$}
{\bf Input:} Integers $i,j,k$\\
{\bf Output:} An index $t^*$ such that $$f(i-1,j-1,t^*)=\max\limits_{0\leq t<r}\left\{f(i-1,j-1,t)|\sigma(P[1:t]\oplus x_i)=k\right\}$$
\begin{algorithmic}[1]
\STATE $tmp \leftarrow -1$, $t^* \leftarrow -1$\\
\FOR{$t=0$ to $r-1$}
\IF{$\sigma(t,x_i)=k \ \AND\  f(i-1,j-1,t)>tmp$}
\STATE $tmp \leftarrow f(i-1,j-1,t), t^* \leftarrow t$\\
\ENDIF
\ENDFOR
\RETURN $t^*$
\end{algorithmic}
\end{algorithm}

Then the value of $u$ in line 10 of Algorithm 1 must be $$u=f(i-1,j-1,t^*)=f(i-1,j-1,\max\sigma(i,j,k)).$$

We can improve the efficiency of above algorithms further in following two points.

First, we can pre-compute a table $\lambda$ of the function $\sigma(P[1:k]\oplus ch)$ for each character $ch\in\sum$ and $1\leq k\leq r$ to speed up the computation of $\max\sigma(i,j,k)$.

\begin{algorithm}
\caption{$\lambda(1:r,ch\in \Sigma)$}
{\bf Input:} String $P=p_1\cdots p_r$, alphabet $\Sigma$\\
{\bf Output:} A table $\lambda$
\begin{algorithmic}[1]
\FORALL{$a\in \Sigma$ \AND \ $a\neq p_1$}
\STATE $\lambda(0,a) \leftarrow 0$\\
\ENDFOR
\STATE $\lambda(0,p_1) \leftarrow 1$\\
\FOR{$t=1$ to $r-1$}
\FORALL{$a\in \Sigma$}
\IF{$a=p_{t+1}$}
\STATE $\lambda(t,a) \leftarrow t+1$\\
\ELSE
\STATE $\lambda(t,a) \leftarrow \lambda(kmp(t),a)$\\
\ENDIF
\ENDFOR
\ENDFOR
\end{algorithmic}
\end{algorithm}

The time cost of above preprocessing algorithm is obviously $O(r|\Sigma|)$. By using this pre-computed table $\lambda$, the value of function $\sigma(P[1:k]\oplus ch)$ for each character $ch\in\sum$ and $1\leq k<r$ can be computed readily in $O(1)$ time.

Second, the computation of function $\max\sigma(i,j,k)$ is very time consuming and many repeated computations are overlapped in the whole {\bf for} loop of the Algorithm 1. We can amortized the computation of function $\max\sigma(i,j,k)$ to each entry of $f(i,j,k)$ in the {\bf for} loop on variable $k$ of the Algorithm 1 and finally reduce the time costs of the whole algorithm.
The modified algorithm can be described as follows.

\begin{algorithm}
\caption{STR-EC-LCS}
{\bf Input:} Strings $X=x_1\cdots x_n$, $Y=y_1\cdots y_m$ of lengths $n$ and $m$, respectively, and a constraint string $P=p_1\cdots p_r$ of lengths $r$\\
{\bf Output:} The length of an LCS of $X$ and $Y$ excluding $P$ as a substring
\begin{algorithmic}[1]
\FORALL{$i,j,k$ , $0\leq i\leq n, 0\leq j\leq m$, and $0\leq k \leq r$}
\STATE $f(i,0,k) \leftarrow 0, f(0,j,k) \leftarrow 0$ \{boundary condition\}
\ENDFOR
\FOR{$i=1$ to $n$}
\FOR{$j=1$ to $m$}
\FOR{$k=0$ to $r$}
\STATE $f(i,j,k) \leftarrow \max\{f(i-1,j,k),f(i,j-1,k)\}$
\ENDFOR
\IF {$x_i=y_j$}
\FOR{$k=0$ to $r$}
\STATE $t \leftarrow \lambda(k,x_i)$
\STATE $f(i,j,t) \leftarrow \max\{f(i,j,t),1+f(i-1,j-1,k)\}$
\ENDFOR
\ENDIF
\ENDFOR
\ENDFOR
\RETURN $\max\limits_{0\leq t<r}\{f(n,m,t)\}$
\end{algorithmic}
\end{algorithm}

Since $\lambda(k,x_i)$ can be computed in $O(1)$ time for each $x_i,1\leq i\leq n$ and any $0\leq k<r$, the loop body of above algorithm requires only $O(1)$ time. Therefore, our new algorithm for computing the length of an LCS of $X$ and $Y$ excluding $P$ as a substring requires $O(nmr)$ time and $O(r|\Sigma|)$ preprocessing time.

If we want to get the answer LCS of $X$ and $Y$ excluding $P$ as a substring, but not just its length, we can also present a simple recursive back tracing algorithm for this purpose as the following Algorithm 7.

\begin{algorithm}
\caption{$back(i,j,k)$}
{\bf Comments:} A recursive back tracing algorithm to construct the answer LCS
\begin{algorithmic}[1]
\IF{$i=0 \ \OR \ j=0$}
\RETURN
\ENDIF
\IF {$x_i=y_j$}
\IF {$f(i,j,k)=f(i-1,j-1,k)$}
\STATE $back(i-1,j-1,k)$
\ELSE
\STATE $back(i-1,j-1,\max\sigma(i,j,k))$
\PRINT $x_i$
\ENDIF
\ELSIF{$f(i-1,j,k)>f(i,j-1,k)$}
\STATE $back(i-1,j,k)$
\ELSE
\STATE $back(i,j-1,k)$
\ENDIF
\end{algorithmic}
\end{algorithm}

In the end of our new algorithm, we will find an index $t$ such that $f(n,m,t)$ gives the length of an LCS of $X$ and $Y$ excluding $P$ as a substring. Then, a function call $back(n,m,t)$ will produce the answer LCS accordingly.

Since the cost of the algorithm $\max\sigma(i,j,k))$ is $O(r)$ in the worst case, the algorithm $back(i,j,k)$ will cost $O(r\max(n,m))$.

Finally we summarize our results in the following Theorem.

\begin{theo}\label{th2}
The Algorithm 6 solves STR-EC-LCS problem correctly in $O(nmr)$ time and $O(nmr)$ space, with preprocessing time $O(r|\Sigma|)$.
\end{theo}

\section{Concluding Remarks}
We have suggested a new dynamic programming solution for the STR-EC-LCS problem. The new algorithm corrects a previously presented dynamic programming algorithm with the same time and space complexities.

The STR-IC-LCS problem is another interesting generalized constrained longest common subsequence (GC-LCS) which is very similar to the STR-EC-LCS problem.

The STR-IC-LCS problem , introduced in\cite{1}, is to find an LCS of two main sequences, in which a constraining sequence of length $r$ must be included as its substring. In \cite{1} an $O(nmr)$-time algorithm was given for it. Almost immediately the presented algorithm was improved to a quadratic-time algorithm and furthermore to many main input sequences\cite{4}.

It is not clear that whether the same improvement can be applied to our presented $O(nmr)$-time algorithm for the STR-EC-LCS problem to achieve a quadratic-time algorithm. We will investigate the problem further.

\end{document}